\documentclass[twocolumn,showpacs,preprintnumbers,amsmath,amssymb]{revtex4}

\def\bftau{\mbox{\boldmath $\tau$}}

\newcommand{\be}{\begin{equation}}
\newcommand{\ee}{\end{equation}}
\newcommand{\bea}{\begin{eqnarray}}
\newcommand{\eea}{\end{eqnarray}}
\newcommand{\ba}{\begin{array}}
\newcommand{\ea}{\end{array}}

\def\bbox{{\,\lower0.9pt\vbox{\hrule \hbox{\vrule height 0.2 cm
\hskip 0.2 cm \vrule height 0.2 cm}\hrule}\,}}
\newcommand{\dsl}{\pa \kern-0.5em /}

\usepackage{graphicx}
\usepackage{dcolumn}
\usepackage{bm}

\begin{document}

\preprint{DAMTP-2006-18, ITFA-2006-08, hep-th/0602260}

\title{Hidden supersymmetry of domain walls and  cosmologies}

\author{Kostas Skenderis}
\affiliation{%
Institute for Theoretical Physics, University of Amsterdam,\\
Valckenierstraat 65, 1018 XE Amsterdam, The Netherlands.
}%

\author{Paul K. Townsend}
\affiliation{
Department of Applied Mathematics and
Theoretical Physics,\\
Centre for Mathematical Sciences, University of Cambridge,\\
Wilberforce Road, Cambridge, CB3 0WA, U.K.
}%


\begin{abstract}

We show that all domain-wall solutions of gravity coupled to 
scalar fields for which the worldvolume 
geometry is Minkowski or anti-de Sitter admit Killing spinors, 
and satisfy corresponding first-order equations involving a 
superpotential  determined by the solution. By analytic continuation, 
all flat or closed FLRW cosmologies are shown to satisfy similar
first-order equations arising from the existence of 
``pseudo-Killing'' spinors. 

\end{abstract}

\pacs{04.65.+e\ \ 11.27.+d\ \  98.80.Jk}
\maketitle

Scalar fields arise naturally in many supergravity theories and the 
domain wall solutions they allow are important for the holographic 
interpretation of renormalization group flow. They have also been 
invoked in the context of inflationary cosmology, and as a possible 
source of `dark energy'.  A general theoretical framework for these 
studies, in spacetime dimension $D=d+1$,  is provided by the 
Lagrangian density
\be\label{Lstart}
{\cal L} = \sqrt{-\det g}\left[ R
  -\frac{1}{2}\left|\partial\Phi\right|^2 
-V(\Phi)\right]
\ee
for metric $g$, with scalar curvature $R$, and scalar fields 
$\Phi$ taking values in some Riemannian target space, on which 
there is some potential energy function $V$.  One purpose of this 
paper is to exploit a close connection between domain wall and 
cosmological solutions of the above model, but our initial 
focus will be domain walls because a domain wall solution may be 
a supersymmetric solution of some supergravity 
theory for which (\ref{Lstart}) is a consistent truncation. 
In practice, this involves a determination of whether the solution
admits Killing spinors, which are non-zero spinor fields 
annihilated by a covariant derivative operator constructed 
from the standard spin connection and a `superpotential', which
determines the potential $V$ by a simple formula involving first derivatives.

There are many  `flat'  domain wall solutions, with Minkowski $d$-dimensional 
geometry, that have long been known to be supersymmetric 
solutions of  some supergravity theory. More  recently, beginning with  
\cite{Behrndt:2001mx},  supersymmetric {\it curved} domain wall solutions 
have been found. 
Such results all depend on the specific superpotential 
of the supergravity theory under study, even though only the potential $V$
is relevant to the solution itself. Moreover,  {\it the superpotential is not 
uniquely defined by the potential}, which means  that there can be
many supergravity theories with the {\it same}  metric-scalar truncation; 
a solution that is non-symmetric for one supergravity theory could be 
a supersymmetric solution of another one.
This state of affairs suggests a supergravity-independent definition of a 
`supersymmetric'  domain wall solution
as one that admits a Killing spinor for {\it some} superpotential function that 
yields the given potential $V$ \cite{Skenderis:1999mm,DeWolfe:1999cp,
Freedman:2003ax,Celi:2004st,Sonner:2005sj}; the superpotential then
defines a ``fake supergravity''  \cite{Freedman:2003ax}. This definition raises 
two related questions:  which domain wall solutions of a given model, with specified 
target space and potential $V$,  are `supersymmetric'  in the above sense, 
and which models admit  such solutions? 

These questions have been raised and partially answered  in the recent 
literature \cite{Freedman:2003ax,Celi:2004st,Sonner:2005sj}.
Here we present 
an essentially complete answer to both questions for domain wall 
solutions that are foliated (or `sliced') by 
$d$-dimensional de Sitter (dS), Minkowski or anti-de Sitter (adS) 
spaces. Moreover, the answer is very simple, and model-independent. 
{\it All} Minkowski and adS sliced domain walls are supersymmetric 
for {\it any} model of the form (\ref{Lstart}), and the only dS-sliced 
domain walls that are supersymmetric are the dS-foliation of $D$-dimensional 
Minkowski or adS space. 
There {\it are} some caveats, in particular 
the result may be true only locally (in the many scalar case) or 
`piecewise' (if the superpotential turns out to be multi-valued). 

Although cosmologies cannot be supersymmetric (with the exception 
of anti-de Sitter space), first-order equations for flat cosmologies arise
in  the Hamilton-Jacobi formalism \cite{Liddle:2000cg} and their similarity
with `BPS'  equations has been noted  \cite{Bazeia:2005tj}. In addition, many 
previous works have obtained cosmological solutions of {\it particular} models
by analytic continuation of domain wall solutions. Here we establish a general 
result: for every domain wall solution (of the type specified above) there is a corresponding 
FLRW cosmology, of the same model but with opposite sign potential.  
The generality of our result for domain walls then implies that {\it all} flat or 
closed FLRW cosmologies solve first-order `BPS-type' equations involving a 
superpotential determined by the solution, despite the fact that they 
are not supersymmetric! We will show that this result arises from the
existence of ``pseudo-Killing'' spinors. 

As we wish to consider  both domain walls and cosmologies, it is 
convenient to introduce a sign $\eta$ such that $\eta=1$ for domain 
walls and $\eta=-1$ for cosmologies. Then, in either case,  
the metric for the solutions of interest here can be put in the form
\be\label{ansatz} 
ds^2_D = \eta \left(fe^{\alpha\varphi}\right)^2 dz^2 + 
e^{2\beta\varphi}\left[
- \frac{\eta\, dr^2}{1+ \eta k r^2} + r^2 d\Omega_\eta^2\right]
\ee
where  $d\Omega_\eta^2$ is {\it either} (for $\eta=-1$) a unit 
radius $SO(d)$-invariant metric on the $(d-1)$-sphere, {\it or} 
(for $\eta=1$) a unit radius $SO(1,d-1)$-invariant metric on the 
$(d-1)$-dimensional hyperboloid. For $\eta=-1$, $z$ is a time 
variable and the  constant $z$ hypersurfaces are  maximally symmetric 
spaces with inverse radius of curvature $k$ normalized to $k=-1,0,1$. 
{}For $\eta=1$, $r$ is the time variable and the
constant $z$ hypersurfaces are maximally symmetric 
$d$-dimensional spacetimes  with inverse radius of curvature $k=-1,0,1$, and 
hence have adS, Minkowski, or dS geometry, respectively. In suitable 
coordinates for the metric $d\Omega_\eta^2$, the $D$-dimensional 
domain-wall and FLRW cosmology  metrics are  related by a 
double-analytic continuation. 

We have allowed for an arbitrary function $f(z)$ in the ansatz 
(\ref{ansatz}) in order to maintain $z$-parametrization invariance, 
and we have introduced for later convenience the $D$-dependent
constants 

\be
\alpha = (D-1) \beta\, , \ \qquad 
\beta =1/\sqrt{2(D-1)(D-2)}\, . 
\ee
The scalar fields must be taken to be functions only of $z$ in order 
to preserve the spacetime isometries.  The field equations then reduce 
to equations for the variables $(\varphi,\Phi)$ that are equivalent 
to the Euler-Lagrange  equations of the effective Lagrangian
\be
L= \frac{\eta}{2} f^{-1} \left(\dot\varphi^2 -
|\dot\Phi|^2\right) 
- fe^{2\alpha\varphi}\left(V
- \frac{k}{2 \beta^2} e^{-2\beta\varphi}\right) \, , 
\ee
where the overdot indicates differentiation with respect to $z$. Note 
that a change in the sign $\eta$ can be compensated by a change of 
sign of both $V$ and $k$, because the overall sign of $L$ has 
no effect on the equations of motion. It follows that  {\it for 
every domain wall solution of a model with scalar potential $V$ 
there is a corresponding FLRW cosmology, with the opposite sign of 
$k$ if $k\ne0$, for a model with the opposite sign of $V$}. 

For simplicity of presentation, we begin by supposing that that there 
is only one field $\sigma$. Later we will show how our results 
generalize to the multi-scalar case. We will also fix the
$z$-reparametrization by the gauge choice $f= e^{-\alpha\varphi}$. In
this gauge, and for a one-scalar model, the Euler-Lagrange equations 
of $L$ are equivalent to the equations
\be \label{feq}
\ddot\varphi = -\alpha\dot\sigma^2 -\left(k\eta/\beta\right)\, 
e^{-2\beta\varphi}\, ,\qquad
\ddot\sigma = -\alpha \dot\varphi\dot\sigma + \eta V'\, , 
\ee
where the prime indicates differentiation with respect to $\sigma$, 
together with the constraint
\be\label{Friedman}
\dot\varphi^2 -\dot\sigma^2 =-2\eta\left[V- 
\frac{k}{2\beta^2}\, e^{-2\beta\varphi}\right]\, . 
\ee
If $V$ has an extremum that allows a solution with $\dot\sigma\equiv0$ 
then the domain wall or cosmological solution is actually a
dS, Minkowski or adS vacuum solution. So we shall 
assume that $\dot\sigma$ is not identically zero. In fact, we shall 
assume initially that $\dot\sigma$ is nowhere zero, returning
subsequently to consider what happens when $\dot\sigma$ has isolated 
zeros. Given that $\dot\sigma\ne0$, there is an inverse function 
$z(\sigma)$ that allows any function of $z$ to be considered as a 
function of $\sigma$. In particular,  given any solution with 
$\eta k \le0$ for which $\dot\sigma\ne0$, we may define a 
complex function 
\be
Z(\sigma) = \omega(\sigma) e^{i\theta(\sigma)}
\ee
by the formulae
\bea\label{om}
\omega &=& \frac{1}{2\alpha} \sqrt{\dot\varphi^2 - 
\frac{k\eta}{\beta^2}e^{-2\beta\varphi}}\, , \\
\theta' &=&  \pm  \sqrt{-k\eta} \,  \left(\frac{\alpha}{\beta}\right)
\dot\sigma\, e^{-\beta\varphi}
\left(\dot\varphi^2 -\frac{k\eta}{\beta^2}
e^{-2\beta\varphi}\right)^{-1}
\label{thetaprime}
\eea
Note that $\theta'=0$ for $k=0$ so in this case we may choose 
$\theta=0$, and hence $Z=\omega$. 

We claim that the function $Z(\sigma)$ constructed according to the 
above prescription satisfies 
\be\label{VZ}
V= 2 \eta \left[|Z'|^2 -\alpha^2|Z|^2\right]\, 
\ee
as a consequence of the equations of motion, and further that the 
solution used to construct $Z(\sigma)$ satisfies,
\bea\label{firstorder}
&&\dot\sigma = \pm 2|Z'| \, ,
\qquad  \dot\varphi = \mp \frac{2\alpha}{ |Z'|} \,  
{\cal R}e \left(\bar Z Z'\right) \, ,\nonumber \\
&& -k\eta\, e^{2\beta \varphi} = \left(2 \alpha \beta\, {\cal I}m\, 
\left(\bar Z Z'\right)/ |Z'|\right)^2\, . 
\eea
In fact, these equations imply the second-order ones.
Inserting (\ref{firstorder}) in (\ref{VZ}) yields the constraint
(\ref{Friedman}). Differentiating the first of (\ref{firstorder}) 
and using (\ref{VZ}) and (\ref{firstorder}) yields the second of 
eqs. (\ref{feq}). Finally, the first of eqs. (\ref{feq}) follows 
directly from the second of eqs. (\ref{firstorder})
upon using the definitions of $\omega, \theta$ in 
(\ref{om})-(\ref{thetaprime}). There is a consistency condition 
between the second and third of eqs. (\ref{firstorder}):
$\dot{\varphi}$ computed from the third 
should agree with the second. This requires
\be\label{Zone}
{\cal I}m\, \left[ \bar Z' \left(Z'{}' + 
\alpha\beta Z\right)\right] =0\, . 
\ee
Remarkably,  this  is an {\it identity}  for $(\omega,\theta)$ 
defined by (\ref{om})-(\ref{thetaprime}), so all $k\eta\le 0$
solutions satisfy first-order equations, for either choice of the sign $\eta$!
 
As a concrete illustration of the above, consider the $D=3$ model 
with $V = -\eta$, and the $k =-\eta$ solution \cite{Sonner:2005sj}
\be
e^\varphi = 1 + e^{\sqrt{2}\,  z}\, ,\qquad e^{-\sigma} = 
1 + e^{-\sqrt{2}\, z}
\ee
For $\eta=1$ this yields an adS-sliced ``separatrix wall '' solution 
that interpolates between an $adS_2\times R$ linear-dilaton vacuum 
(at $z=-\infty$) and the $adS_3$ vacuum (at $z=\infty$). For $\eta=-1$ 
it yields a $k=1$ FLRW cosmology that interpolates between an Einstein 
Static Universe (supported by a constant $\sigma$ kinetic energy) in 
the far past and the $dS_3$ vacuum in the far future. Note that 
$\sigma<0$ for this solution, so that $(1-e^\sigma)$ is positive. One finds
that
\bea
\omega(\sigma) &=& \sqrt{1-e^\sigma +
\frac{1}{2} e^{2 \sigma}}\, , \\
\theta(\sigma) &=& \arctan \left[e^{-\sigma}
\sqrt{2\left(1-e^\sigma\right)}\right]  \nonumber\\
&&+\ \frac{1}{\sqrt{2}} \log\left(\frac{1-\sqrt{1-e^\sigma}}
{1+ \sqrt{1-e^\sigma}}\right) +
\theta_0, \nonumber
\eea
for arbitrary, and irrelevant, constant $\theta_0$. 

So far, we have considered domain walls and cosmologies on 
an equal footing, but we now restrict to the domain wall case, 
$\eta=1$. For this case, we claim that the first-order equations 
(\ref{firstorder}) are `BPS' equations that guarantee the existence 
of a Killing spinor field. It would be sufficient for our purposes to
consider a complex superpotential modeled on minimal $D=4$ 
supergravity but to make use of previous work on, or inspired by,
minimal $D=5$ supergravity we consider instead a real $Sp_1$-triplet 
superpotential ${\bf W}(\sigma)$ and a Killing spinor 
equation of the form \cite{LopesCardoso:2001rt,Freedman:2003ax,Celi:2004st}
\be\label{KS}
\left(D_\mu -\alpha\beta {\bf W}\cdot \bftau\, 
\Gamma_\mu\right)\epsilon=0\, , \quad (\mu=0,1,\dots,d),
\ee
where $D_\mu$ is the standard covariant derivative on spinors, and 
$\bftau$ is the triplet of Pauli matrices. In the context of minimal $D=5$
supergravity, $\epsilon$ is an $Sp_1$-Majorana spinor and
${\bf W}$ is real. The reality of ${\bf W}$ is 
also required for the  ``gamma-trace'' of the Killing spinor equation to be a 
Dirac equation with a hermitian ``mass''  matrix, and this condition can (and should) 
be imposed as part of the definition of a `fake'  Killing spinor. 
With this understood, we may allow $\epsilon$ in (\ref{KS}) 
to be a Dirac spinor in arbitrary spacetime dimension $D$.

{}For a solution of the assumed form, the Killing spinor equation (\ref{KS})
reduces to the equations
\bea\label{killing}
\partial_z \epsilon &=& \alpha\beta \, {\bf W}\cdot \bftau \, 
\Gamma_{\underline z} \, \epsilon \, ,\\
\hat D_m \epsilon&=& e^{\beta\varphi}\hat\Gamma_m\left[
\left(\beta/2\right)\, \dot\varphi   \, 
\Gamma_{\underline z} + \alpha\beta\, 
{\bf W} \cdot\bftau \right]\epsilon\, , \nonumber 
\eea
where $\Gamma_{\underline z}$ is a {\it constant} matrix that squares to the identity,
and a hat indicates restriction to the (normalized) 
worldvolume metric, so $\hat\Gamma_m$ are the worldvolume Dirac 
matrices. The integrability conditions of these
equations were discussed in \cite{Freedman:2003ax} and we review this
analysis here.
The second of the equations (\ref{killing}) has the integrability condition
\be\label{integrability1}
\dot\varphi^2 = 4\alpha^2 |{\bf W}|^2 + 
(k/\beta^2)\, e^{-2\beta\varphi}\, . 
\ee
We will now suppose that the potential $V$ is given in terms of 
the triplet superpotential by the relation
\be\label{VW}
V= 2\left[|{\bf W}'|^2 -\alpha^2 |{\bf W}|^2\right]\, . 
\ee
At this point, the reader may guess how ${\bf W}$ is determined 
by the complex function $Z$ introduced earlier, but no guesswork 
is needed: the relation between the two will emerge from consistency 
requirements. Given the constraint (\ref{Friedman}) and the above 
form of the potential, (\ref{integrability1}) implies that
\be\label{firstsig}
\dot\sigma = \pm 2|{\bf W}'|\, .
\ee
Differentiating (\ref{integrability1})  and using the equations of 
motion to eliminate $\ddot\varphi$, and then eliminating $V$ in 
favour of ${\bf W}$, we deduce the  first-order equation 
\be\label{firstphi}
\dot\varphi = \mp 2\alpha 
\left({\bf W}\cdot {\bf W}' \right)/ |{\bf W}'|
\ee
{\it and} the condition
\be\label{kcond}
|{\bf W}\times {\bf W}'|^2 = -k \left(D-2\right)^2 
e^{-2\beta\varphi}\, |{\bf W}'|^2\, .
\ee
It follows from this condition that a dS-sliced ($k=1$) domain-wall  
can admit Killing spinors only if ${\bf W}$ is constant, which 
requires $\dot\sigma\equiv 0$ and implies that $V$ is a non-positive
constant; in this case the D-dimensional spacetime is a $dS$-foliation of
either Minkowski or adS space. Excluding 
these trivial cases, we conclude that a Killing spinor requires either 
$k=0$ or $k=-1$, and that $k=0$ requires 
${\bf W}\times {\bf W}' ={\bf 0}$, which implies
that ${\bf W}= W {\bf n}$ for a singlet superpotential 
$W(\sigma)$ and a {\it fixed} $3$-vector ${\bf n}$.

The Killing spinor equations (\ref{killing}) also have the joint 
integrability condition
\be
\left(\dot\sigma + 2 {\bf W}'\cdot\bftau\, 
\Gamma_{\underline z}\right)\epsilon =0\, ,
\ee
which has the supergravity interpretation as the condition of vanishing 
supersymmetry variation of the superpartner to $\sigma$. This
condition must be satisfied for all $z$; it is trivially satisfied if $\dot\sigma\equiv0$
since (\ref{firstsig}) then implies that ${\bf W}$ is constant. Otherwise, 
it  implies the projection
\be\label{proj}
\left(1\pm \Gamma\right) \epsilon =0 \, ,\qquad 
\Gamma = \left({\bf W}'/|{\bf W}'|\right)\cdot\bftau\, 
\Gamma_{\underline z}\, .
\ee
For $k=0$ we have $\Gamma= ({\bf n}\cdot\bftau) \Gamma_{\underline z}$, 
which is a constant traceless matrix that squares to the identity 
matrix, implying preservation of 1/2 supersymmetry. Otherwise 
$\Gamma$ is not a constant matrix and differentiation with respect 
to $z$ of the projection condition yields the consistency condition 
\be\label{consist}
\left({\bf W}'{}' + \alpha\beta {\bf W} \right) \times {\bf W}' = {\bf 0}\, . 
\ee
This condition implies that ${\bf W}$ and all its derivatives are 
coplanar, so that ${\bf W} = X{\bf n} + Y{\bf m}$ for fixed
orthonormal $3$-vectors ${\bf n}$ and ${\bf m}$, and functions 
$X(\sigma)$, $Y(\sigma)$.  The first-order equations (\ref{firstsig}) 
and (\ref{firstphi}) are then equivalent  to the first-order equations 
(\ref{firstorder}) if we make the identification $Z= X+iY$. 
The integrability condition (\ref{integrability1})  is then equivalent to 
the equation (\ref{om}) for $\omega(\sigma)$, and 
(\ref{kcond}) is similarly equivalent  to (\ref{firstorder}), 
which is itself equivalent to the equation (\ref{thetaprime}) for
$\theta(\sigma)$. 

Thus, the complex function $Z$ appearing in 
(\ref{firstorder}) determines the triplet superpotential ${\bf W}$.
In terms of $Z$, the consistency condition (\ref{consist}) is just
(\ref{Zone}) and, as already mentioned,  this is satisfied
identically! This means that {\it all flat or adS-sliced domain wall 
solutions of the one-scalar model for which $\dot\sigma$ does not 
vanish preserve $1/2$ supersymmetry for a superpotential that is 
determined by the solution, in the sense that they admit Killing
spinors for this superpotential subject to (at most) a $1/2$ 
supersymmetry projection.} 

The condition of non-vanishing $\dot\sigma$ was needed 
because our construction of the
superpotential assumed the existence of a function $z(\sigma)$ inverse 
to $\sigma(z)$. While this condition may be satisfied for many 
domain-wall solutions, others will typically have isolated values 
of $z$ for which $\dot\sigma=0$ (for example, the ``$\lambda$-perturbed Janus
solutions'' described in \cite{Sonner:2005sj} all have one point at which 
$\dot\sigma=0$). When this happens the inverse function $z(\sigma)$ 
will  become multi-valued, with different branches in intervals of 
$z$ on either side of a zero of $\dot\sigma(z)$. In other words, it 
will still be true that the domain wall solution is supersymmetric 
for a superpotential determined by the solution, but this 
superpotential will be a multi-valued function and more than 
one branch will be needed. Thus understood, our claim 
remains true  `piecewise' even when $\dot\sigma(z)$ has 
zeros. 

So far we have restricted our analysis to single-scalar models, and 
at first sight it might seem unlikely that the main result could 
generalize to models with an arbitrary number of scalars and an 
arbitrary potential for them. However, a simple argument  shows 
that it {\it does} generalize, at least locally. The key observation 
\cite{Celi:2004st} is that for any domain-wall solution, the functions 
$\Phi(z)$ define a curve in the scalar field target space, and this 
curve may be chosen as one of the  `axes'  of  a new set of
curvilinear coordinates on the target space, in which case, the 
equations defining the curve state that all scalar fields  but one,
call it $\sigma$,  are constant. On this curve the potential is a function 
only of $\sigma$ and the problem is thus reduced to the one already solved, 
except of course that the change of target space coordinates needed to 
achieve this may not be valid globally. However, our result, that {\it all flat or 
adS-sliced domain wall solutions are supersymmetric} remains true locally. 

Although our Killing spinor results were derived assuming real triplet superpotential,  
inspired by $D=5$ supergravity, they are valid for any $D$. We could have obtained these
results by considering a simpler Killing spinor equation with a complex superpotential, 
such as one would find in $D=4$ by dimensional reduction of the $D=5$ case (although
the discussion to follow on cosmology would then be more involved). 
No new possibilities can arise from considering more general superpotentials (as 
confirmed by the results of \cite{Zagermann:2004ac} for a 5-vector superpotential)
because novelty for our purposes would require $k=1$ and there is no physically acceptable 
supersymmetric extension of the $dS$ isometry algebra. It is therefore satisfying 
that, with the exception of the $dS$-foliations of Minkowski or $adS$ (for which the isometry 
algebra is enlarged),  we have not found any supersymmetric  $k=1$ domain walls (although 
this does not preclude the possibility of first-order equations  \cite{Afonso:2006gi}). 

We conclude with some comments on the cosmology case. Recall that any 
domain wall solution has an associated cosmological solution with flipped signs 
of $V$ and $k$. At first it appears that such solutions cannot have Killing spinors
because $\Gamma_{\underline z}$ now squares  to minus the identity, so (\ref{proj}) 
has no non-zero solutions. However, we must also take ${\bf W}\to i{\bf W}$ in order
to flip the signs of $V$ and $k$, as is clear from (\ref{VW}) and  (\ref{kcond}).
We now have what appears to be a Killing spinor for any $k\ge0$ cosmological solution,
but the ${\bf W}\to i{\bf W}$ step replaces the initial hermitian ``mass''  matrix 
${\bf W}\cdot\bftau$ in the gamma-traced Killing spinor equation
by an anti-hermitian one. As explained earlier, this means that we no longer 
have a {\it bona fide} Killing spinor, although we 
 do have what might be called a ``pseudo-Killing'' spinor. It is unclear what the 
implications of the existence of pseudo-Killing spinors are, but
 their existence nevertheless explains why $k\ge0$ FLRW cosmologies 
are also driven by first order equations. The implications of this 
fact remain to be explored.

\begin{acknowledgments}
KS is supported by NWO. PKT thanks the EPSRC for a Senior Research Fellowship. 
\end{acknowledgments}

\end{document}